\documentclass[conference,a4paper]{IEEEtran}

\usepackage[utf8]{inputenc} 
\usepackage[T1]{fontenc}
\usepackage{url}              

\usepackage[cmex10]{amsmath}  
\usepackage{mleftright}       
\mleftright                   

\usepackage{graphicx}         
\usepackage{booktabs}         

\usepackage[greek,english]{babel}
\usepackage{teubner}

\usepackage[numbers, compress]{natbib}
\usepackage[inline]{enumitem}
\usepackage{amsfonts,mathtools,centernot}
\usepackage{dsfont}
\usepackage{textcomp}
\usepackage{dirtytalk}
\usepackage{color, soul}
\usepackage{balance}
\usepackage{nicefrac}

\def\BibTeX{{\rm B\kern-.05em{\sc i\kern-.025em b}\kern-.08em
    T\kern-.1667em\lower.7ex\hbox{E}\kern-.125emX}}

\DeclarePairedDelimiter\abs{\lvert}{\rvert}%

\DeclareMathOperator*{\esssup}{ess\,sup}
\newcommand{\nll}{\centernot{\ll}}

\newtheorem{theorem}{Theorem}
\newtheorem{corollary}{Corollary}
\newtheorem{lemma}{Lemma}
\newtheorem{definition}{Definition}

\newtheorem{remark}{Remark}
\newtheorem{example}{Example}

\newcounter{relctr} 
\everydisplay\expandafter{\the\everydisplay\setcounter{relctr}{0}} 

\AtBeginDocument{} 

\allowdisplaybreaks

\IEEEoverridecommandlockouts
\begin{document}

\title{Pointwise Maximal Leakage on General Alphabets} 

\author{%
  \IEEEauthorblockN{Sara~Saeidian\IEEEauthorrefmark{1},
                    Giulia~Cervia\IEEEauthorrefmark{2},
                    Tobias~J.~Oechtering\IEEEauthorrefmark{1},
                    and Mikael Skoglund\IEEEauthorrefmark{1}}
  \IEEEauthorblockA{\IEEEauthorrefmark{1}%
                    KTH Royal Institute of Technology,
                    100 44 Stockholm, Sweden,
                    \{saeidian, oech, skoglund\}@kth.se}
  \IEEEauthorblockA{\IEEEauthorrefmark{2}%
                    IMT Nord Europe,
                    Centre for Digital Systems,
                    F-59000 Lille,
                    France,
                    giulia.cervia@imt-nord-europe.fr}
\thanks{This work has been supported by the Strategic Research Agenda Program, Information and Communication Technology – The Next Generation (SRA ICT – TNG) funded by the Swedish Government and KTH Digital Futures center within the project DataLEASH.}
}

\maketitle
\everypar{\looseness=-1}
\begin{abstract}
Pointwise maximal leakage (PML) is an operationally meaningful privacy measure that quantifies the amount of information leaking about a secret $X$ to a single outcome of a related random variable $Y$. In this paper, we extend the notion of PML to random variables on arbitrary probability spaces. We develop two new definitions: First, we extend PML to countably infinite random variables by considering adversaries who aim to guess the value of discrete (finite or countably infinite) functions of $X$. Then, we consider adversaries who construct estimates of $X$ that maximize the expected value of their corresponding gain functions. We use this latter setup to introduce a highly versatile form of PML that captures many scenarios of practical interest whose definition requires no assumptions about the underlying probability spaces.
\end{abstract}

\section{Introduction}
\label{sec:intro}


Recently, the problem of private data analysis has attracted much attention from an information-theoretic perspective. A wide variety of privacy measures, for example, mutual information~\cite{asoodeh2015maximal, wang2016relation, makhdoumi2014information, rassouli2021perfect, liao2019tunable}, measures based on $f$-divergences~\cite{diaz2019robustness, rassouli2019optimal}, probability of correctly guessing~\cite{asoodeh2018estimation}, information privacy~\cite{jiang2021context,du2012privacy}, and log-lift~\cite{hsu2019information, sadeghi2021properties} are studied that aim to quantify the amount of information leaking about a (private) random variable $X$ by disclosing a related random variable $Y$. (See~\cite{bloch2021overview} for a recent survey on information-theoretic privacy measures.) \emph{Pointwise maximal leakage} (PML)~\cite{saeidian2022pointwise, saeidian2022pointwise_isit} is one such measure that is particularly robust and operationally meaningful. Introduced by~\citet{saeidian2022pointwise}, PML is a generalization of the pre-existing notion of \emph{maximal leakage}~\cite{smith2009foundations, braun2009quantitative,issa2019operational}. While maximal leakage quantifies the information leaking from the average outcome of the random variable $Y$, PML can be used to measure the information leaking from each individual outcome $Y=y$. Hence, the pointwise definition sets up a more flexible framework in which information leakage is viewed as a random variable that can be bounded and controlled in different ways~\cite[Sec. III]{saeidian2022pointwise}. The original definition of maximal leakage is also retrieved as the expected value of the information leakage random variable. 

To define PML,~\cite{saeidian2022pointwise} exploits and unifies two seemingly different formalizations of maximal leakage: the \emph{randomized function view}~\cite{issa2019operational} and the \emph{gain function view}~\cite{alvim2012measuring}. The randomized function view of leakage assumes that an adversary attempts to guess the value of an arbitrary discrete (randomized) function of $X$, denoted by $U$. Then, PML is defined as (the logarithm of) the multiplicative increase in the probability of correctly guessing the value of $U$ after observing an outcome $Y=y$, compared to the prior probability of correctly guessing the value of $U$. On the other hand, the gain function view of leakage considers an adversary who wishes to maximize the expected value of an arbitrary non-negative gain function. In this case, PML is defined as (the logarithm) of the multiplicative increase in the expected gain of an adversary who has observed $Y=y$, compared to the prior expected gain. In~\cite[Thm. 2]{saeidian2022pointwise}, it is shown that when $X$ takes values in a finite alphabet, the two definitions of leakage are in fact equivalent.

While PML is a robust and meaningful privacy measure, currently its definition is restricted to finite alphabet random variables. Our goal in this paper is to extend the notion of PML to random variables on general measurable spaces. In~\cite[Thm. 7]{issa2019operational}, \citeauthor{issa2019operational} undertake a similar task and define a version of maximal leakage on general alphabets. Their definition, however, has certain limitations. Most importantly, while $X$ is assumed to be a random variable on a measurable space, their setup still concerns an adversary who aims to guess the value of a finite alphabet random variable $U$, i.e., they consider the randomized function view of leakage. While this setup is suitable when $X$ has a discrete (finite or countably infinite) alphabet, it results in a conceptually weaker definition in the general case as it is assumed that adversaries do not exploit the generalized structure of the space of $X$ which is no longer restricted to be purely atomic. For example, if $X$ is an absolutely continuous random variable (with respect to the Lebesgue measure), then an adversary can reasonably aim to construct a real-valued guess of $X$ so as to maximize a gain function that is decreasing in the mean squared error. Such attack scenarios cannot be directly studied using the randomized function view of leakage. 

In this paper, we generalize the definition of PML in two directions: First, in Section~\ref{sec:discrete_def}, we extend the randomized function view of leakage to countable probability spaces (Thm.~\ref{thm:countable_thm}). Then, in Section~\ref{sec:general_def}, we use the gain function view of leakage to obtain a universal definition of PML that requires no assumptions about the underlying probability spaces (Thm.~\ref{thm:general_leakage}). This latter setup considers an adversary who is interested in maximizing the expected value of an arbitrary non-negative and measurable gain function, so the resulting notion of privacy is highly robust. We show that in both cases PML can be written as the Rényi divergence~\cite{renyi1961measures, van2014renyi} of order infinity of the posterior distribution of $X$ from the prior distribution of $X$.  Finally, in Corollary~\ref{cor:privacy_rv} we discuss the nuances of defining an information leakage random variable in the general setup and show how densities can be useful for deriving a form of PML that is a measurable function of $Y$. Throughout the paper, we give examples of common attack scenarios and evaluate PML for typical mechanisms, for example, adding independent Gaussian noise to a Gaussian random variable $X$ (Example~\ref{ex:gaus_1}).

\section{Notation and Background}
\label{sec:background}
\vspace{-0.1em}
\subsection{Notation}
\vspace{-0.1em}
We adopt the following notational conventions: $\mathbb R_+ = [0,\infty)$, $\bar{\mathbb R}_+ = [0,\infty]$, $\mathbb Z = \{\ldots, -1, 0, 1, \ldots\}$, $\mathbb N=\{0,1, \ldots\}$, $\mathbb N^* = \{1, 2, \ldots\}$, and $[n] = \{1, \ldots, n\}$ with $n \in \mathbb  N^*$. Sets are represented by uppercase letters, e.g., $E$, and $\mathds 1_E$ denotes the indicator function of $E$. Collections of sets are represented by calligraphic letters, e.g., $\mathcal E$. If $E$ is a topological space then $\mathcal B_E$ denotes the Borel $\sigma$-algebra on $E$. Given a measurable space $(E, \mathcal E)$, we use $\mathcal E_+$ to denote the set of all functions that are measurable relative to $\mathcal E$ and $\mathcal B_{\bar{\mathbb R}_+}$. Suppose $\mu$ and $\nu$ are measures on $(E, \mathcal E)$ and assume that $\mu$ is $\sigma$-finite. If $\nu \ll \mu$, then we write $p = \frac{d\nu}{d\mu}$, or alternatively, $\nu(dx) = p(x) \, \mu(dx)$ to imply that $\int_E f(x) \; \nu(dx) = \int_E f(x) \, p(x) \; \mu(dx)$ for all $f \in \mathcal E_+$, where $p \in \mathcal E_+$ is the Radon-Nikodym derivative of $\nu$ with respect to $\mu$. 

Throughout the paper, we assume that a probability space $(\Omega, \mathcal H, \mathbb P)$ is fixed in the background. Given $f \in \mathcal H_+$, the essential supremum of $f$ with respect to $\mathbb P$ is $\esssup_{\mathbb P} f = \sup \{c \in \mathbb R_+ \colon \mathbb P(f > c) > 0\}$. A mapping $X : \Omega \to E$ is called a random variable taking values in $(E, \mathcal E)$ if $X$ is measurable relative to $\mathcal H$ and $\mathcal E$. We use $P_X$ to denote the distribution of $X$. A mapping $P_{Y \mid X} : E \times \mathcal F \to \mathbb [0,1]$ is called a transition probability kernel (or simply kernel) from $(E, \mathcal E)$ into $(F, \mathcal F)$ if the mapping $x \mapsto P_{Y \mid X=x}(B)$ is in $\mathcal E_+$ for all $B \in \mathcal F$, and $P_{Y \mid X=x}(\cdot)$ is a probability measure on $(F, \mathcal F)$ for all $x \in E$. Let $P_{XY}$ be a probability measure on the product space $(E \times F, \mathcal E \otimes \mathcal F)$ with marginals $P_X$ and $P_Y$. Then we write $P_{XY} (dx, dy) = P_X(dx) \, P_{Y \mid X=x}(dy)$ to imply that $\mathbb E \, f = \int_{E \times F} f(x,y) P_{XY} (dx,dy) = \int_E P_X(dx) \int_F f(x,y) \; P_{Y \mid X=x} (dy)$ for all $f \in (\mathcal E \otimes \mathcal F)_+$. 

Suppose $Y$ is a random variable taking values in $(F, \mathcal F)$. Let $\sigma Y$ denote the $\sigma$-algebra generated by $Y$ on $\Omega$. We use $\mathbb E[f \mid Y]$ to denote the conditional expectation of $f \in \mathcal H_+$ given $\sigma Y$. Since $\mathbb E[f \mid Y] \in (\sigma Y)_+$, then there exists $\phi \in \mathcal F_+$ such that $\mathbb E[f \mid Y] = \phi \circ Y$. Hence, we use the notation $\mathbb E[f \mid Y=y]$ to represent $\phi(y)$ for each $y \in F$. 
\vspace{-0.2em}
\subsection{PML for finite alphabet random variables}
We begin by recalling the definition of Rényi divergence of order infinity~\cite{renyi1961measures, van2014renyi}, which we later use to provide simplified expressions for PML.  
\begin{definition}[{Rényi divergence of order $\infty$~\cite[Thm. 6]{van2014renyi}}] 
Let $P$ and $Q$ be probability measures on the measurable space $(\Omega, \mathcal H)$. Let $\mu$ be a $\sigma$-finite measure satisfying $P \ll \mu$ and $Q \ll \mu$. The Rényi divergence of order $\infty$ of $P$ from $Q$ is defined as
\vspace{-0.5em}
\begin{equation*}
\label{eq:renyi_div_inf}
    D_\infty(P \Vert Q) = \log \sup_{A \in \mathcal H} \frac{P(A)}{Q(A)} = \log \left(\esssup_{P} \frac{p}{q}\right),
\end{equation*}
where $p = \frac{dP}{d\mu}$ and $q = \frac{dQ}{d\mu}$. 
\end{definition}
If $P \ll Q$ the divergence can also be expressed as 
\begin{align*}
    D_\infty(P \Vert Q) = \log \left(\esssup_{P} \frac{dP}{dQ} \right) = \log \left(\esssup_{Q} \frac{dP}{dQ} \right).
\end{align*}
On the other hand, if $P \nll Q$, then $D_\infty(P \Vert Q) = \infty$.\footnote{We use the conventions that $\nicefrac{0}{0} = 1$ and $\nicefrac{x}{0} = \infty$ if $x > 0$.} When the sample space $\Omega$ is countable we may write~\eqref{eq:renyi_div_inf} in the form
\begin{equation*}
    D_\infty(P \Vert Q) = \log \left(\sup_{\omega \in \Omega} \frac{P(\omega)}{Q(\omega)} \right).
\end{equation*}

In~\cite{saeidian2022pointwise}, \citeauthor{saeidian2022pointwise} use two different threat models to introduce PML on finite spaces: the randomized function view of leakage~\cite[Def. 1]{saeidian2022pointwise}  based on the setup of~\citet{issa2019operational}, and the gain function view of leakage~\cite[Cor. 1]{saeidian2022pointwise} based on the setup of~\citet{alvim2012measuring}. The following definition is a generalization of~\cite[Def. 1]{saeidian2022pointwise} where the alphabets of $X$, $Y$, and $U$ are no longer restricted to be finite but are allowed to be countable.

\begin{definition}[Randomized function view of leakage]
\label{def:randomized_function_view}
Suppose $X$ is a random variable on the discrete set $E$, and $Y$ is a random variable on the discrete set $F$. Let $P_{XY}$ denote the joint distribution of $X$ and $Y$. The pointwise maximal leakage from $X$ to $y \in F$ is defined as\footnote{To be able to define the leakage for all $y \in F$ when $F$ is a countable set, we may assume that $\mathbb P(\cdot \mid Y=y) = \mathbb P(\cdot)$ if $P_Y(y) = 0$. That is, conditioning on events with probability zero equals no conditioning.}
\vspace{-0.4em}
\begin{equation*}
    \ell_{P_{XY}}(X\to y) \coloneqq \log \sup_{P_{U \mid X}} \frac{\sup_{P_{\hat U \mid Y}} \mathbb P \left(U=\hat U \mid Y=y \right)}{\max_{u \in G} P_U(u)},
\end{equation*}
\vspace{-0.2em}
where $U$ is any random variable on a countable set $G$ such that the Markov chain $U-X-Y$ holds.
\end{definition}

\begin{theorem}[{\cite[Thm. 1]{saeidian2022pointwise}}]
\label{thm:previous_Sara}
If $X$ has a finite support, then the pointwise maximal leakage from $X$ to $y \in F$, described by Definition~\ref{def:randomized_function_view}, is given by
\vspace{-0.2em}
\begin{equation*}
    \ell_{P_{XY}}(X \to y) = D_\infty(P_{X\mid Y=y} \Vert P_X),
\end{equation*}
\vspace{-0.2em}
where $P_{X \mid Y=y}$ denotes the posterior distribution of $X$ given $y \in F$. 
\end{theorem}

\vspace{-0.4em}
\section{PML for Discrete Random Variables: Randomized Function View}
\label{sec:discrete_def}
In this section, we show that the result of Theorem~\ref{thm:previous_Sara} can be extended to cases where the support of $X$ is countably infinite. Note that the randomized function view described by Definition~\ref{def:randomized_function_view} is a special case of the more general gain function view presented in Definition~\ref{def:general_leakage} (see Example~\ref{ex:randomized_function_view}). Still, Definition~\ref{def:randomized_function_view} is of independent interest for us since it provides a strong operational meaning for PML on discrete probability spaces while its corresponding result in Theorem~\ref{thm:countable_thm} can be proved without any measure theoretic intricacies. 
\begin{theorem}
\label{thm:countable_thm}
Let $X$ and $Y$ be random variables taking values in the countable sets $E$ and $F$, respectively. The pointwise maximal leakage from $X$ to $y \in F$, described by Definition~\ref{def:randomized_function_view}, is given by
\vspace{-0.2em}
\begin{equation*}
    \ell_{P_{XY}}(X\to y) = D_\infty(P_{X \mid Y=y} \Vert P_X).
\end{equation*}
\end{theorem}
\vspace{-0.2em}
Theorem~\ref{thm:countable_thm} is proved in Appendix~\ref{ssec:proof_countable_thm}.

\begin{example}[Geometric distribution]
Suppose $X \sim \mathrm{Geom} (p)$ with $p \in (0,1)$. Let $Y$ be a binary random variable defined through the kernel $P_{Y \mid X=x} (0) = 1 - P_{Y \mid X=x} (1) = q^{x}$ with $q \in (0,1)$ and $x \in \mathbb N^*$. Then, $P_Y(0) = 1- P_Y(1) = \frac{pq}{1 - q + pq}$, and 
\vspace{-1em}
\begin{gather*}
    \ell_{P_{XY}}(X \to 0) = \log \; \sup_{x \in \mathbb N^*} \; \frac{P_{Y \mid X=x}(0)}{P_Y(0)} = \log \; \frac{1 - q + pq}{p},\\
    \ell_{P_{XY}}(X \to 1) = \log \; \sup_{x \in \mathbb N^*} \; \frac{P_{Y \mid X=x}(1)}{P_Y(1)} = \log \; \frac{1 - q + pq}{1-q}.\\
\end{gather*}
\end{example}

\vspace{-2em}
\section{PML on Arbitrary Probability Spaces: Gain Function View}
\label{sec:general_def}
In this section, we present a universal definition of PML that not only showcases the robustness of PML as a privacy measure but also requires no assumptions about the underlying probability spaces. The setup is an extension of~\cite{alvim2012measuring}. 
\begin{definition}[Gain function view of leakage]
\label{def:general_leakage}
Suppose $X$ is a random variable taking values in $(E, \mathcal E)$ with distribution $P_X$ and $Y$ is a random variable taking values in $(F, \mathcal F)$. Let $P_{XY}$ denote the joint distribution of $X$ and $Y$. The pointwise maximal leakage from $X$ to $y \in F$ is 
\begin{equation}
\label{eq:general_leakage}
    \ell_{P_{XY}}(X \to y) \coloneqq \log \sup_{\substack{(D, \mathcal D), \\ g \in \Gamma}} \frac{\sup\limits_{P_{W \mid Y}} \; \mathbb E \left[g(X, W) \mid Y=y \right]}{\sup_{w \in D} \; \mathbb E \left[g(X,w) \right]},
\end{equation}
where the supremum in the numerator is over all transition probability kernels $P_{W \mid Y}$ from $(F, \mathcal F)$ into $(D, \mathcal D)$, and $\Gamma$ denotes the set of all gain functions defined as 
\begin{equation*}
    \Gamma \coloneqq \{g \in (\mathcal E \otimes \mathcal D)_+ \mid \sup_{w \in D} \mathbb E \left[g(X,w) \right] < \infty \}.
\end{equation*}
\end{definition}
The above definition models an adversary who is interested in constructing an estimate of $X$, denoted by $W$, which would maximize the expected value of her gain function. $W$ is a random variable taking values in $(D, \mathcal D)$ and gain functions are picked from the set $\Gamma$. Then, to measure the amount of information leaking about $X$ by disclosing an outcome $Y=y$, we evaluate the ratio of the adversary's expected gain given observation $y \in F$, and her expected gain without any observations. PML is then defined by taking the supremum of this ratio over all possible measurable spaces $(D, \mathcal D)$ and all $g \in \Gamma$. Note that the requirement $\sup_{w \in D} \mathbb E \left[g(X,w) \right] < \infty$ implies that the adversary chooses a function such that her expected gain can potentially be improved upon observing $y \in F$. 

Below, we provide two examples of gain functions that describe typical attack scenarios. The first one concerns an adversary who wishes to guess the value of a discrete function of $X$, denoted by $U$, which retrieves the setup of~\cite[Thm. 7]{issa2019operational}. This setup can be used to model a hypothesis-testing adversary. For example, we may take $U = \mathds{1}_{A^*} (X)$ to model a binary hypothesis test for determining whether or not $X$ is in the set $A^* \in \mathcal E$. The second example describes an adversary who aims to approximate the value of a random variable on a separable metric space. 
\begin{example}[Guessing a discrete function of $X$]
\label{ex:randomized_function_view}
Suppose $U$ is a discrete random variable taking values in the set $A$ and induced by the kernel $P_{U \mid X}$. To model an adversary who is interested in guessing the value of $U$ we let $D = A$, define $\mathcal D$ to be the discrete $\sigma$-algebra on $A$ (i.e., its power set), and express the gain function $g_\bullet$ as follows: 
\vspace{-0.3em}
\begin{equation*}
    g_\bullet (x,w) = P_{U \mid X=x} (w), \quad x \in E, w \in D. 
\end{equation*}
In this case, the denominator of~\eqref{eq:general_leakage} describes the prior probability of correctly guessing the value of $U$ whereas the numerator represents the posterior probability of correctly guessing $U$ given $Y=y$. 
\end{example}


\begin{example}[Approximate guessing in metric spaces]
Let $(A,\rho)$ be a complete and separable metric space. Suppose $U$ is a random variable taking values in $(A, \mathcal B_A)$ induced by a kernel $P_{U \mid X}$. Fix $\varepsilon > 0$. Our goal is to model an adversary who attempts to guess the value of $U$ within a radius of $\varepsilon$. Suppose $D$ is a countable dense subset of $A$ and $\mathcal D$ is the discrete $\sigma$-algebra on $D$. Let $B_\varepsilon(w) = \{a \in A : \rho(a, w) < \varepsilon\}$ denote the open ball of radius $\varepsilon$ centered at $w \in D$. Consider the gain function $g_\sim$ defined as 
\vspace{-0.3em}
\begin{equation*}
    g_\sim(x,w) = P_{U \mid X=x}(B_\varepsilon(w)), \quad x \in E, w \in D. 
\end{equation*}
Note that the countability of $D$ ensures that $g_\sim$ defined above is $\mathcal E \otimes \mathcal D$-measurable. Then, for fixed $w \in D$ we have 
\begin{align*}
    \mathbb E[g_\sim(X,w)] &= \int P_{U \mid X=x}(B_\varepsilon(w)) \, P_X(dx) = P_U(B_\varepsilon(w))\\
    &= \mathbb P[U \in B_\varepsilon(w)]. 
\end{align*}
Hence, evaluating the denominator of~\eqref{eq:general_leakage} with $g_\sim$ yields the prior probability of approximately guessing $U$. Similarly, it can be verified that the numerator of~\eqref{eq:general_leakage} evaluated with $g_\sim$ describes the posterior probability of approximately guessing $U$ given $Y=y$. 
\end{example}

We are now in place to state the main result of our paper: that PML in the form described by Definition~\ref{def:general_leakage} can too be written as the Rényi divergence of the posterior distribution of $X$ from the prior distribution of $X$. The proof is inspired by a result of~\citet[Thm. 2]{van2014renyi} where it is shown that the general expression for Rényi divergence can be written as the supremum of the divergence evaluated over all finite and measurable partitions of the underlying $\sigma$-algebra. 

Theorem~\ref{thm:general_leakage} requires a single assumption: that the joint distribution $P_{XY}$ can be disintegrated into the marginal $P_Y$ and a transition probability kernel $P_{X \mid Y}$ from $(F, \mathcal F)$ into $(E, \mathcal E)$. We discuss this assumption in Remark~\ref{rem:disintegration}.
\begin{theorem}
\label{thm:general_leakage}
Suppose there exists a transition probability kernel $P_{X \mid Y}$ from $(F, \mathcal F)$ into $(E, \mathcal E)$ such that $P_{XY}(dx, dy) = P_Y(dy) P_{X \mid Y=y}(dx)$. Then, the pointwise maximal leakage from $X$ to $y \in F$, described by Definition~\ref{def:general_leakage}, is given by
\begin{align}
    \ell_{P_{XY}}(X \to y) &= D_\infty(P_{X \mid Y=y} \Vert P_X).
\end{align}
\end{theorem}

\begin{IEEEproof}
Fix $y \in F$. To simplify the numerator of~\eqref{eq:general_leakage} we make use of the following lemma proved in Appendix~\ref{ssec:proof_lemma_numerator}.  
\begin{lemma}
\label{lemma:numerator}
Let $W$ be a random variable induced by a transition probability kernel $P_{W \mid Y}$ from $(F, \mathcal F)$ into $(D, \mathcal D)$. Given $g \in \Gamma$, if there exists a transition probability kernel $P_{X \mid Y}$ from $(F, \mathcal F)$ into $(E, \mathcal E)$ such that $P_{XY}(dx, dy) = P_Y(dy) P_{X \mid Y=y}(dx)$, then
\begin{equation}
\label{eq:leakage_numerator}
    \sup_{P_{W \mid Y}} \mathbb E \left[g(X, W) \!\mid \!Y=y \right] = \sup_{w \in D} \int_{E} g(x,w) P_{X \mid Y=y}(dx),  
\end{equation}
where the supremum on the LHS is over all kernels from $(F, \mathcal F)$ into $(D, \mathcal D)$. 
\end{lemma}

First, we assume that $P_{X \mid Y=y} \ll P_X$. For notational convenience, let $f(x) \coloneqq \frac{dP_{X \mid Y=y}}{dP_X}(x)$ denote the Radon-Nikodym derivative of $P_{X \mid Y=y}$ with respect to $P_X$. We begin by showing that $\ell_{P_{XY}}(X \to y) \leq D_\infty(P_{X \mid Y=y} \Vert P_X)$. Fix an arbitrary measurable space $(D, \mathcal D)$, and a gain function $g \in \Gamma$. We can write 
\begin{align*}
    \frac{\sup\limits_{P_{W \mid Y}} \mathbb E \left[g(X, W) \mid Y=y \right]}{\sup\limits_{w \in D} \mathbb E \left[g(X,w) \right]} &= \frac{\sup\limits_{w \in D} \; \int_{E} g(x,w) \; P_{X \mid Y=y}(dx)}{\sup\limits_{w \in D} \int_E g(x,w) \; P_X(dx)} \\
    &\leq \sup_{w \in D} \frac{\int_{E} g(x,w) \; P_{X \mid Y=y}(dx)}{\int_E g(x,w) \; P_X(dx)}  \\
    &= \sup_{w \in D} \frac{\int_{E} g(x,w) \; f(x) \; P_X(dx)}{\int_E g(x,w) \; P_X(dx)}  \\
    &\leq \esssup_{P_X} f \\
    &= \exp \Big(D_\infty(P_{X \mid Y=y} \Vert P_X) \Big). 
\end{align*}
Thus, $\ell_{P_{XY}}(X \to y) \leq D_\infty(P_{X \mid Y=y} \Vert P_X)$. 

Now, we show that $\ell_{P_{XY}}(X \to y) \geq D_\infty(P_{X \mid Y=y} \Vert P_X)$. Since $P_{X \mid Y=y} \ll P_X$ we may, without loss of generality, assume that $f(x) < \infty$ for all $x \in E$. Let $D = \mathbb Z \cup \{-\infty\}$, and suppose $\mathcal D$ is the discrete $\sigma$-algebra on $D$. Fix $\varepsilon > 0$ and consider the following (countable and disjoint) partition of $E$: 
\begin{equation}
\label{eq:parition}
    B_w^\varepsilon = \{x \in E : e^{w \varepsilon} \leq f(x) < e^{(w+1) \varepsilon}\},\quad w \in D,
\end{equation}
which is indexed by $D$. Note that since $f(x)$ is $\mathcal E$-measurable, $B_w^\varepsilon \in \mathcal E$ for all $w \in D$. Let us define the gain function $g^* : E \times D \to \mathbb R_+$ as follows: 
\begin{equation*}
    g^*(x,w) = \begin{cases}
        \frac{1}{P_X(B_w^\varepsilon)} \, \mathds{1}_{B_w^\varepsilon} (x) & \mathrm{if} \; P_X(B_w^\varepsilon) > 0,\\
        0 & \mathrm{if} \; P_X(B_w^\varepsilon) = 0. 
    \end{cases}
\end{equation*}
Then, we can write 
\begin{subequations}
\begin{align}
    &\exp \Big(\ell_{P_{XY}}(X \to y) \Big) \geq \frac{\sup_{P_{W \mid Y}} \mathbb E \left[g^*(X, W) \mid Y=y \right]}{\sup_{w \in D} \mathbb E \left[g^*(X,w) \right]} \nonumber \\
    &= \frac{\sup_{w \in D} \; \int_{E} g^*(x,w) \; P_{X \mid Y=y}(dx)}{\sup_{w \in D} \int_E g^*(x,w) \; P_X(dx)} \nonumber \\
    &= \frac{\sup\limits_{w \in D : P_X(B_w^\varepsilon) > 0} \; \int_{E} \frac{1}{P_X(B_w^\varepsilon)} \, \mathds{1}_{B_w^\varepsilon} (x) \; P_{X \mid Y=y}(dx)}{\sup\limits_{w \in D : P_X(B_w^\varepsilon) > 0} \int_E \frac{1}{P_X(B_w^\varepsilon)} \, \mathds{1}_{B_w^\varepsilon} (x) \; P_X(dx)}\nonumber \\
    &= \frac{\sup\limits_{w \in D : P_X(B_w^\varepsilon) > 0} \frac{P_{X \mid Y=y} (B_w^\varepsilon)}{P_X(B_w^\varepsilon)}}{\sup\limits_{w \in D : P_X(B_w^\varepsilon) > 0} \frac{P_X (B_w^\varepsilon)}{P_X(B_w^\varepsilon)}} \nonumber\\
    &= \sup\limits_{w \in D : P_X(B_w^\varepsilon) > 0} \frac{P_{X \mid Y=y} (B_w^\varepsilon)}{P_X(B_w^\varepsilon)} \label{subeq:achiev_1} \\
    &= \esssup_{P_X} \; \bar{f}, \label{subeq:achiev_2}
\end{align}
\end{subequations}
where
\vspace{-1.2em}
\begin{equation*}
    \bar{f}(x) \coloneqq \sum_{w \in D} \frac{P_{X \mid Y=y}(B_w^\varepsilon)}{P_X(B_w^\varepsilon)} \; \mathds{1}_{B_w^\varepsilon} (x), \quad x \in E.
\end{equation*}
In~\eqref{subeq:achiev_2}, we have written~\eqref{subeq:achiev_1} as a function of $x$. We have replaced the supremum over $w$ in~\eqref{subeq:achiev_1} with the essential supremum over $x$ in~\eqref{subeq:achiev_2} because $\bar{f}$ is constant on each set $B_w^\varepsilon$. Note that $\bar{f}(x) < \infty$ for all $x \in E$ even if there exists $w \in D$ such that $P_X(B_w^\varepsilon) = 0$. This is because $P_{X \mid Y=y} \ll P_X$ and we use the convention that $\nicefrac{0}{0} = 1$. 

Let $\mathcal G \coloneqq \sigma \{B_w^\varepsilon\}$ denote the $\sigma$-algebra on $E$ generated by the collection of sets $\{B_w^\varepsilon\}$. We argue that $\bar{f}$ is (a version of) the conditional expectation of $f$ given $\mathcal G$, that is, $\bar{f} = \mathbb E \left[f \mid \mathcal G \right]$. Clearly, $\bar{f}$ is $\mathcal G$-measurable, so we should verify that $\int_A f \; dP_X = \int_A \bar{f} \; dP_X$ for all $A \in \mathcal G$. It is, however, sufficient to verify this equality for $A = B_w^\varepsilon$ because each non-empty set in $\mathcal G$ can be written as a countable union of sets in $\{B_w^\varepsilon\}$ and the monotone convergence theorem ensures that $\int_{\cup_i C_i} f \; dP_X = \sum_i \int_{C_i} f \; dP_X$ for each countable collection of disjoint sets $\{C_i\}$ in $\mathcal G$. Thus, by noting that 
\vspace{-0.5em}
\begin{equation*}
    \int_{B_w^\varepsilon} \bar{f} \; dP_X = P_{X \mid Y=y} (B_w^\varepsilon) = \int_{B_w^\varepsilon} f \; dP_X,
\end{equation*}
for all $w \in D$ we conclude that $\bar{f} = \mathbb E \left[f \mid \mathcal G \right]$. 

Finally, we can write 
\begin{subequations}
\begin{align}
    \ell_{P_{XY}}(X \to y) &\geq \log \; \esssup_{P_X} \; \mathbb E \left[f \mid \mathcal G \right] \nonumber\\
    &\geq \log \left( \left(\esssup_{P_X} f\right) \; e^{-\varepsilon} \right) \label{subeq:achiev_3}\\
    &= \log \; \esssup_{P_X} f - \varepsilon \nonumber\\
    &= D_\infty(P_{X \mid Y=y} \Vert P_X) - \varepsilon,\nonumber
\end{align}
\end{subequations}
where~\eqref{subeq:achiev_3} is due to the fact that by the definition of the sets $\{B_w^\varepsilon\}$ in~\eqref{eq:parition}, $\mathbb E[f \mid \mathcal G]$ never differs from $f$ by more than a factor of $e^\varepsilon$. Then, letting $\varepsilon \to 0$, we obtain $\ell(X \to y) \geq D_\infty(P_{X \mid Y=y} \Vert P_X)$, which completes the proof for the case $P_{X \mid Y=y} \ll P_X$.

On the other hand, if $P_{X \mid Y=y} \nll P_X$ then there exists $A_0 \in \mathcal E$ such that $P_X(A_0) = 0$ and $P_{X \mid Y=y}(A_0) >0$. Let $(D, \mathcal D)$ be an arbitrary measurable space, and consider the gain function $g(x,w) = \mathds{1}_{A_0} (x)$ for all $w \in D$. Then, it is easy to see that $\mathbb E \left[g(X, W) \mid Y=y \right] = P_{X \mid Y=y}(A_0) >0$ for all kernels $P_{W \mid Y}$ while $\sup_{w \in D} \; \mathbb E \left[g(X,w) \right] = 0$. Hence, $\ell(X \to y) = D_\infty(P_{X \mid Y=y} \Vert P_X) = \infty$, as desired.  
\end{IEEEproof}

\begin{remark}
\label{rem:disintegration}
Theorem~\ref{thm:general_leakage} assumes that the joint distribution $P_{XY}$ can be disintegrated into the marginal $P_Y$ and a kernel $P_{X \mid Y}$. This can be achieved in different ways. For example, we may start with a distribution $P_Y$ and a kernel $P_{X \mid Y}$ and construct $P_{XY}$ such that it satisfies $P_{XY} (dx, dy) = P_Y(dy) P_{X \mid Y=y} (dx)$. Otherwise, we may assume that $(E, \mathcal E)$ is a Borel space~\cite[Def. 8.35]{klenke2013probability} in which case the existence of a regular version of the conditional probability $\mathbb P [ \; \cdot \mid Y]$ restricted to $\sigma X \subset \mathcal H$ is guaranteed~\cite[Thm. IV.2.18]{ccinlar2011probability}. In this latter case, $P_{X \mid Y}$ is any kernel satisfying $\mathbb P [X \in A \mid Y] (\omega) = P_{X \mid Y=Y(\omega)} (A)$ for $\mathbb P$-almost all $\omega \in \Omega$ and all $A \in \mathcal E$. Hence, Theorem~\ref{thm:general_leakage} requires that $P_{XY}$ can be disintegrated into a marginal distribution and a kernel, though it is immaterial how this is actually achieved. For a detailed discussion on disintegration theorems and the existence of regular conditional probabilities see~\cite{faden1985existence}.
\end{remark}

Equipped with Theorem~\ref{thm:general_leakage}, we can calculate the information leaking from an observation $y \in F$. However, this result alone is insufficient for obtaining an information leakage random variable $\ell_{P_{XY}}(X \to Y)$. The difficulty is that the mapping $y \mapsto \ell_{P_{XY}}(X \to y)$ must be $\mathcal F$-measurable and there are certain nuances associated with this task. For example, we need to ensure that if $P_{X \mid Y=y} \ll P_X$, then the Radon-Nikodym derivative $\frac{dP_{X \mid Y=y}}{dP_X}$ is jointly measurable in $(x,y)$, or that the set $\{y \in F : P_{X \mid Y=y} \ll P_X \}$ is measurable. 

To obtain a measurable version of $\ell_{P_{XY}}(X \to y)$ we use the pragmatic assumption that the joint distribution $P_{XY}$ is absolutely continuous with respect to the product of two $\sigma$-finite measures on $(E, \mathcal E)$ and $(F, \mathcal F)$~\cite[Sec. 2.6]{polyanskiy2014lecture}. This assumption also has the advantage of guaranteeing that $P_{XY}(dx, dy) = P_Y(dy) P_{X \mid Y=y}(dx)$ holds. 

\begin{corollary}[Privacy leakage random variable]
\label{cor:privacy_rv}
Suppose $P_{XY}$ is a probability measure on the product space $(E \times F, \mathcal E \otimes \mathcal F)$ satisfying
\vspace{-0.4em}
\begin{equation}
\label{eq:disintegration}
    P_{XY}(dx, dy) = p(x,y) \, \mu(dx) \, \nu(dy), \quad x \in E, y \in F,
\end{equation}
where $\mu$ and $\nu$ are $\sigma$-finite measures on $(E, \mathcal E)$ and $(F, \mathcal F)$ respectively, and $p \in (\mathcal E \otimes \mathcal F)_+$. Then, there exists a transition probability kernel $P_{X \mid Y}$ from $(F, \mathcal F)$ into $(E, \mathcal E)$ such that 
\begin{enumerate}
    \item $P_{XY} (dx, dy) = P_Y(dy) \, P_{X \mid Y=y} (dx)$;
    \item $P_{X \mid Y=y} \ll P_X$ for $\nu$-almost all $y \in F$; and
    \item the mapping $y \mapsto \ell_{P_{XY}}(X \to y)$ is $\mathcal F$-measurable. 
\end{enumerate}
\end{corollary}
Corollary~\ref{cor:privacy_rv} is proved in Appendix~\ref{ssec:proof_cor_measurability}.

\begin{remark}
The assumption of Corollary~\ref{cor:privacy_rv} guarantees that $P_{XY} \ll P_X \otimes P_Y$, where $P_{X} \otimes P_Y$ denotes the product of $P_X$ and $P_Y$. This assumption also allows us to write PML in different forms using densities: 
\vspace{-0.3em}
\begin{align*}
    \ell_{P_{XY}}(X \to y) &= \esssup_{P_X} i(X;y)\\
    &=\log \left(\esssup_{P_X} \frac{f_{X \mid Y} (X,y)}{f_X(X)}\right)\\
    &= \log \left(\esssup_{P_X} \frac{f_{Y \mid X}(y,X)}{f_Y(y)}\right),
\end{align*}
where $i(X;Y) = \log \frac{dP_{XY}}{dP_X \otimes P_Y} (X,Y)$ denotes the information density (the expectation of which is mutual information), $f_{X \mid Y} \in (\mathcal E \otimes \mathcal F)_+$ denotes the density of $P_{X \mid Y}$ with respect to $\mu$, and $f_X \in \mathcal E_+$ denotes the density of $P_X$ with respect to $\mu$. Densities $f_{Y \mid X}$ and $f_Y$ are defined similarly. 
\end{remark}

Below, we calculate PML when $X$ has Gaussian distribution and $Y$ is obtained by adding independent Gaussian noise to $X$. Further examples are provided in Appendix~\ref{ssec:examples}. Here, densities are defined with respect to the Lebesgue measure. 
\begin{example}[Additive Gaussian Noise]
\label{ex:gaus_1}
Suppose $Y = X + N$ where $X \sim \mathcal N(0, \sigma_X^2)$, $N \sim \mathcal N(0, \sigma_N^2)$, and $X$ and $N$ are independent. The PML from $X$ to $y \in \mathbb R$ is given by 
\vspace{-0.7em}
\begin{align*}
    \ell_{P_{XY}}(X \to y) &= \log \; \sup_{x \in \mathbb R} \frac{f_{Y \mid X}(y,x)}{f_Y(y)}\\
    &= \frac{1}{2} \log \left( 1 + \frac{\sigma_X^2}{\sigma_N^2} \right) + \frac{y^2}{2 (\sigma_X^2 + \sigma_N^2)}. 
\end{align*}
As expected, for fixed $y \in \mathbb R$ and $\sigma_X^2$, taking $\sigma_N^2 \to \infty$ implies $\ell_{P_{XY}}(X \to y) \to 0$. 
\end{example}

\vspace{-1em}
\section{Conclusions}
\label{sec:conc}
In this paper, we have extended the notion of PML to random variables with general alphabets. Theorem~\ref{thm:countable_thm} describes a direct extension of the randomized function view of leakage from finite to countably infinite random variables. On the other hand, Theorem~\ref{thm:general_leakage} illustrates that the gain function view of leakage can be used to define PML on arbitrary probability spaces. 

The following points are worth emphasizing: 

\textbf{Properties of PML.} We have shown that in all cases PML can be written as the Rényi divergence of order $\infty$ of the posterior distribution of $X$ from the prior distribution of $X$. Hence, some properties of PML, such as non-negativity and satisfying a data-processing inequality, are directly inherited from the Rényi divergence~\cite{van2014renyi}. We leave the detailed development of the properties of PML as future work. 

\textbf{Mechanism design via PML.} In Examples~\ref{ex:gaus_1} and~\ref{ex:gaus_2}, we have calculated PML in two typical setups involving the Gaussian distribution. These examples signify the advantage of PML over the average-case notion of maximal leakage: In many situations, including these Gaussian examples and even the discrete case of  Example~\ref{ex:poisson}, $\exp \Big(\ell_{P_{XY}}(X \to Y)\Big)$ is not $P_Y$-integrable (i.e., its expectation is infinite). However, one can still characterize privacy using probability bounds on PML. Such bounds can be useful for designing privacy-preserving mechanisms, where simple mechanisms are conceived by adding noise with suitable parameters to $X$. We shall explore mechanism design with PML in future works.

\bibliographystyle{IEEEtranN}
\footnotesize
\bibliography{IEEEabrv, main}
\balance
\normalsize
\newpage
\appendix
\subsection{Proof of Theorem~\ref{thm:countable_thm}}
\label{ssec:proof_countable_thm}
The finite case has been proved in~\cite{saeidian2022pointwise}, so here we assume that $E = \{x_1, x_2, \ldots \}$ is countably infinite. Fix $y \in F$. The argument for showing that $\ell_{P_{XY}}(X \to y) \leq D_\infty(P_{X \mid Y=y} \Vert P_X)$ is identical to the finite case laid out in~\cite[Thm. 1]{saeidian2022pointwise}, so we only prove that $\ell_{P_{XY}}(X \to y) \geq D_\infty(P_{X \mid Y=y} \Vert P_X)$. Fix a small constant $\delta > 0$ and let $k$ be an integer satisfying $\sum_{k}^ \infty P_X(x_k) = P_X(A_\delta) < \delta$, where $A_\delta \coloneqq \{x_k, x_{k+1}, \ldots\}$. For notational convenience, we will use the shorthands $\xi(B) \coloneqq \displaystyle{\frac{\sum_{x \in B }P_{X \mid Y=y}(x)}{\sum_{x \in B} P_X(x)}}$ with $B \subset E$, and 
\begin{equation*}
    \ell_U(X \to y) \coloneqq \log \frac{\sup_{P_{\hat U \mid Y}} \mathbb P \left(U=\hat U \mid Y=y \right)}{\max_{u 
    \in G} P_U(u)},
\end{equation*}
for a given random variable $U$ induced by a kernel $P_{U \mid X}$. We prove the inequality for the following two cases: First, we assume that there exists $x^* \in E$ such that $\log \,\xi(x^*) = D_\infty(P_{X \mid Y=y} \Vert P_X)$. Then, we consider the case where the supremum in the expression of the Rényi divergence is not attained by any $x \in E$. 

Suppose $\log \, \xi(x^*) = D_\infty(P_{X \mid Y=y} \Vert P_X)$. Assume $\delta$ is sufficiently small so that $x^*$ is not in the set $A_\delta$. Define the finite random variable $W$ with alphabet $D = \{1, \ldots, k\}$ according to the kernel $P_{W \mid X=x_i}(i) = 1$ for $i \in [k-1]$ and $P_{W \mid X=x_i} (k) = 1$ for $i \geq k$. Then, $P_W(i) = P_X(x_i)$ for $i \in [k-1]$ and $P_W(k) = P_X(A_\delta)$. Now, consider the random variable $U_\text{\stater}$ induced by the \emph{shattering channel} $P_{U_\text{\stater} \mid W}$ defined in~\cite[Thm. 1]{issa2019operational}, which yields 
\begin{subequations}
\begin{align}
    &\ell_{P_{XY}}(X \to y) = \sup_{P_{U \mid X}} \ell_U( X \to y)\nonumber\\
    &\geq \ell_{U_\text{\stater}} (X \to y)\nonumber\\
    &= \log \; \max_{i \in [k]} \; \frac{P_{W \mid Y=y} (i)}{P_W(i)}\label{subeq:countable_1}\\ 
    &= \log \; \max \left\{\xi(x_1), \ldots, \xi(x^*), \ldots, \xi(x_{k-1}), \xi(A_\delta) \right \}\nonumber\\
    &\geq \log \; \xi(x^*) = D_\infty(P_{X \mid Y=y} \Vert P_X),\nonumber
\end{align}
\end{subequations}
where~\eqref{subeq:countable_1} is due to the definition of the shattering channel. Hence, $\ell_{P_{XY}}(X \to y) \geq D_\infty(P_{X \mid Y=y} \Vert P_X)$ holds in the first case. 

Next, suppose the supremum in the expression of the Rényi divergence is not attained by any $x \in E$.  Let $M = D_\infty(P_{X \mid Y=y} \Vert P_X)$, where $M \in (0, \infty]$. Since the supremum is not attained by any $x$, then we must have $\limsup\limits_{n \to \infty} \; \log \, \xi(x_n) = M$. Fix an arbitrary $0 < m < M$, and define the set $B_m = \{x \in E\ \colon \xi(x) > m \}$. Clearly, for each $ 0 < m < M$ and $0< \delta <1$ the two sets $A_\delta$ and $B_m$ have non-empty intersection. Define $E_{m, \delta} \coloneqq A_\delta \cap B_m$ and $E_{m, \delta}' \coloneqq  A_\delta \setminus B_m$. Similarly to the previous case, we define a finite random variable $W'$ with alphabet $D' = \{1, \ldots, k+1\}$ according to the kernel $P_{W' \mid X=x_i}(i) = 1$ for $i \in [k-1]$, $P_{W' \mid X=x} (k) = 1$ for \vspace{0.5em} $x \in E_{m, \delta}$, and $P_{W' \mid X=x} (k+1) = 1$ for $x \in E_{m, \delta}'$. Now, consider the random variable $U_\text{\stater}'$ induced by the shattering channel $P_{U_\text{\stater}' \mid W'}$. We obtain
\begin{align*}
    &\ell_{P_{XY}}(X \to y) \geq \ell_{U_\text{\stater}'} (X \to y)\\
    &= \log \; \max_{i \in [k+1]} \; \frac{P_{W' \mid Y=y} (i)}{P_{W'}(i)}\\ 
    &= \log \; \max \left\{\xi(x_1), \ldots, \xi(x_{k-1}), \xi(E_{m,\delta}), \xi(E_{m,\delta}') \right \}\\
    &\geq \log \; \xi(E_{m,\delta}) = \log \; \frac{\sum_{x \in E_{m,\delta}} P_{X \mid Y=y} (x)}{\sum_{x \in E_{m,\delta}} P_X(x)}\\
    &\geq \log \; \inf_{x \in E_{m,\delta}} \frac{P_{X \mid Y=y}(x)}{P_X(x)} \geq m, 
\end{align*}
where the last inequality follows by the definitions of the sets $E_{m, \delta}$ and $B_m$. Finally, taking $m \to M$ yields 
\begin{equation*}
    \ell_{P_{XY}}(X \to y) \geq M = D_\infty(P_{X \mid Y=y} \Vert P_X),
\end{equation*}
as desired. 

\subsection{Proof of Lemma~\ref{lemma:numerator}}
\label{ssec:proof_lemma_numerator}
To show that the LHS of~\eqref{eq:leakage_numerator} lower bounds the RHS, fix an arbitrary kernel $P_{W \mid Y}$. We have 
\begin{align*}
    &\mathbb E \left[g(X, W) \mid Y=y \right] = \int_{E \times D} g(x,w) \; P_{XW \mid Y=y} (dx, dw)\\
    &= \int_{D} \; P_{W \mid Y=y}(dw) \int_{E} g(x,w)\; P_{X \mid Y=y}(dx) \\ 
    &\leq \int_{D} \; P_{W \mid Y=y}(dw) \left(\sup_{w \in D} \int_{E} g(x,w)\; P_{X \mid Y=y}(dx) \right)\\
    &= \sup_{w \in D} \int_{E} g(x,w)\; P_{X \mid Y=y}(dx). 
\end{align*}
Taking the supremum over all kernels $P_{W \mid Y}$ we obtain 
\begin{equation*}
    \sup_{P_{W \mid Y}} \mathbb E \left[g(X, W) \mid Y=y \right] \leq \sup_{w \in D} \; \int_{E} g(x,w) \; P_{X \mid Y=y}(dx).  
\end{equation*}
To show that the LHS of~\eqref{eq:leakage_numerator} upper bounds the RHS, fix an arbitrary \vspace{0.3em}  $a < \sup_{w \in D} \int_{E} g(x,w)\; P_{X \mid Y=y}(dx)$. Then, there exists $w' \in D$ such that $\int_{E} g(x,w')\; P_{X \mid Y=y}(dx) \geq a$. Let $\delta_w$ denote the Dirac measure defined by
\begin{equation*}
    \delta_w (A) = \begin{cases}
        1 & \mathrm{if} \; w \in A,\\
        0 & \mathrm{if} \; w \notin A,\\
    \end{cases} 
\end{equation*}
for each $A \in \mathcal D$. We can write 
\begin{align*}
    \sup_{P_{W \mid Y}} \mathbb E & \left[g(X, W) \mid Y=y \right]\\
    &\geq \int_{E} P_{X \mid Y=y}(dx) \int_{D} g(x,w)\; \delta_{w'}(dw) \\ 
    &= \int_{E} g(x,w')\; P_{X \mid Y=y}(dx)\\
    &\geq a. 
\end{align*}
Then, letting $a \to \sup_{w \in D} \int_{E} g(x,w)\; P_{X \mid Y=y}(dx)$ we obtain the desired inequality.     

\subsection{Proof of Corollary~\ref{cor:privacy_rv}}
\label{ssec:proof_cor_measurability}
Define the functions 
\begin{gather*}
    q(y) \coloneqq \int_E p(x,y) \; \mu(dx), \quad y \in F,\\
    r(x) \coloneqq \int_F p(x,y) \; \nu(dy), \quad x \in E,\\
    k(x,y) \coloneqq \begin{cases}
        \frac{p(x,y)}{q(y)} & \mathrm{if} \; q(y) > 0, \\
        r(x) & \mathrm{if} \; q(y) = 0, \\
    \end{cases}, \quad x \in E, y \in F.
\end{gather*}
Let
\begin{equation*}
    P_{X \mid Y=y} (A) \coloneqq \int_A k(x,y) \; \mu(dx), \quad A \in \mathcal E, y \in F.
\end{equation*}
It can easily be checked that $P_{X \mid Y}$ defined above is a transition probability kernel from $(F, \mathcal F)$ into $(E, \mathcal E)$. 

First, we show that $P_{X \mid Y=y} \ll P_X$ holds $\nu$-almost everywhere. Suppose $A_0 \in \mathcal E$ satisfies $P_X(A_0) = 0$. Noting that $P_X(dx) = r(x) \mu(dx)$, we have
\begin{equation*}
    P_X(A_0) = \int_{A_0} r(x) \; \mu(dx) = \int_E \mathds{1}_{A_0}(x) \, r(x) \; \mu(dx) = 0,
\end{equation*}
i.e., $\mathds{1}_{A_0} (x) \, r(x) = 0$ $\mu$-almost everywhere. In other words,
\begin{equation}
\label{eq:absolute_continuity_helper}
    \mu \left( A_0 \cap \{x \in E : r(x) > 0\} \right) = 0. 
\end{equation}
Now, if $q(y) = 0$, then $P_{X \mid Y=y} (A_0) = P_X(A_0) = 0$ by construction. So, suppose $q(y) > 0$. In this case, we have 
\begin{align*}
    P_{X \mid Y=y} (A_0) &= \frac{1}{q(y)} \int_{A_0} p(x,y) \; \mu(dx)\\
    &=  \frac{1}{q(y)} \Bigg( \int_{A_0 \cap \{r > 0\}} p(x,y) \; \mu(dx)\\
    &\hspace{5em}+ \int_{A_0 \cap \{r = 0\}} p(x,y) \; \mu(dx) \Bigg). 
\end{align*}
The first integral is zero due to~\eqref{eq:absolute_continuity_helper}. Moreover, for each $x \in E$, $r(x) = 0$ implies that $p(x,y) = 0$ $\nu$-almost everywhere; thus, the second integral is also zero $\nu$-almost everywhere. We conclude that $P_{X \mid Y=y} \ll P_X$ for $\nu$-almost all $y \in F$. 

Next, we argue that $P_{XY} (dx, dy) = P_Y(dy) \, P_{X \mid Y=y} (dx)$. Noting that $P_Y(dy) = q(y) \, \nu(dy)$ and $P_{X \mid Y=y} (dx) = k(x,y) \mu(dx)$ we write 
\begin{subequations}
\begin{align}
    P_Y(dy) \, P_{X \mid Y=y} (dx) &= q(y) \, k(x,y) \, \mu(dx) \, \nu(dy) \nonumber \\
    &= \begin{cases}
        p(x,y) \, \mu(dx) \, \nu(dy) & \mathrm{if} \; q(y) > 0,\\
        0 & \mathrm{if} \; q(y) = 0.\\
    \end{cases} \nonumber \\
    &= p(x,y) \, \mu(dx) \, \nu(dy) \label{subeq:almost_everywhere_stuff}\\
    &= P_{XY}(dx, dy), \nonumber
\end{align}
\end{subequations}
where~\eqref{subeq:almost_everywhere_stuff} is due to the fact that for each $y \in F$, $q(y) = 0$ implies $p(x,y) = 0$ $\mu$-almost everywhere and a Radon-Nikodym derivative is specified uniquely up to almost everywhere equivalence.  

Finally, we show that the mapping $y \mapsto \ell_{P_{XY}}(X \to y)$ is $\mathcal F$-measurable. Define the set $B_0 = \{y \in F : \int_{\{r = 0 \}} k(x,y) \, \mu(dx) = 0\}$ which is guaranteed to be in $\mathcal F$ by Fubini's theorem. The leakage $\ell_{P_{XY}}(X \to y)$ can be expressed as 
\begin{equation*}
    \ell_{P_{XY}}(X \to y) = \begin{cases}
    \esssup_{P_X} \left( \frac{k(x,y)}{r(x)} \right) & \mathrm{if} \; y \in B_0, \\
    \infty & \mathrm{if} \; y \notin B_0. 
    \end{cases}
\end{equation*}
Note that $\displaystyle{\frac{k(x,y)}{r(x)}}$, which is an $(\mathcal E \otimes \mathcal F)$-measurable function, is used as the Radon-Nikodym derivative $\frac{dP_{X \mid Y=y}}{dP_X}$. It remains to show that the essential supremum of a jointly measurable function is measurable. We state this in the form of a lemma, proved in Appendix~\ref{ssec:proof_lemma_ess_sup}. 
\begin{lemma}
\label{lemma:ess_sup}
Given measurable spaces $(E, \mathcal E)$ and $(F, \mathcal F)$, suppose $s \in (\mathcal E \otimes \mathcal F)_+$. Let $P_X$ be a probability measure on $(E, \mathcal E)$. Then, the function $t : F \to \bar{\mathbb R}_+$ defined as $t(y) = \esssup_{P_X} \, s(x,y)$ is $\mathcal F$-measurable. 
\end{lemma}

Equipped with Lemma~\ref{lemma:ess_sup}, we conclude that the mapping $y \mapsto \ell_{P_{XY}}(X \to y)$ is $\mathcal F$-measurable, as desired.

\subsection{Proof of Lemma~\ref{lemma:ess_sup}}
\label{ssec:proof_lemma_ess_sup}
To show that $t$ is $\mathcal F$-measurable it suffices to show that the inverse image $t^{-1} (c, \infty]$ is in $\mathcal F$ for each $c \in \mathbb R_+$. Fix an arbitrary $c \in \mathbb R_+$. Given $y \in F$, define the set $C_y = \{x \in E : s(x,y) > c\}$ which is in $\mathcal E$ by the measurability of the mapping $x \mapsto s(x,y)$ for fixed $y \in F$. Now, we write
\begin{align*}
    t^{-1} (c, \infty] &= \{y \in F : t(y) > c \}\\
    &= \{y \in F : P_X(\{x\in E : s(x,y) > c\}) > 0\}\\
    &= \{y \in F : P_X(C_y) > 0\}\\
    &= \left \{y \in F : \left( \int_E \mathds{1}_{C_y}(x) \; P_X(dx) \right) > 0 \right \}.
\end{align*}
The mapping $(x,y) \mapsto \mathds{1}_{C_y} (x)$ is $\mathcal E \otimes \mathcal F$-measurable since $\{(x,y) \in E\ \times F : \mathds{1}_{C_y} (x) = 1 \} = \{(x,y) \in E\ \times F : s(x,y) > c \} \in \mathcal E \otimes \mathcal F$. Then, Fubini's theorem ensures that $y \mapsto \int_E \mathds{1}_{C_y}(x) \; P_X(dx)$ is $\mathcal F$-measurable, which in turn, implies that $t^{-1} (c, \infty]$ belongs to $\mathcal F$.

\subsection{Other examples}
\label{ssec:examples}
\begin{example}[Poisson and Binomial distributions]
\label{ex:poisson}
Suppose $X \sim \mathrm{Pois}(\lambda p)$, where $\lambda > 1$, $p \in (0,1)$, and $\lambda (1 - p) < 1$. Assume $Y$ is defined through the kernel 
\begin{equation*}
    P_{Y \mid X=x}(y) = \begin{cases}
        {\displaystyle \frac{\Big(\lambda(1 - p) \Big)^{y-x} \, e^{-\lambda(1 - p)}}{(y - x)!}} & \mathrm{if} \; y \geq x,\\ 
        0 & \mathrm{if} \; y < x, 
    \end{cases}
\end{equation*}
where $x \in \mathbb N$. It can be easily verified that $X \mid Y=y \sim \mathrm{Binom} (y,p)$. Hence, the PML from $X$ to $y \in \mathbb N$ is given by 
\begin{align*}
    \ell_{P_{XY}}(X \to y) &= \log \; \sup_{x \in \mathbb N} \, \frac{P_{X \mid Y=y} (x)}{P_X(x)}\\
    &= \log \; \max\limits_{x \in \{0, \ldots, y \}} \, \frac{P_{X \mid Y=y} (x)}{P_X(x)}\\
    &= \log \; \left( e^{\lambda p} \, \lambda^{-y} \, y! \right). 
\end{align*}
\end{example}

\begin{example}[Gaussian mixtures]
Suppose $X \sim \mathrm{Ber}(\frac{1}{2})$ is an equiprobable Bernoulli random variable, and $Y \mid X=x \sim \mathcal N(x, \sigma^2)$ has Gaussian distribution with mean $x \in \{0,1\}$ and variance $\sigma^2$. The PML from $X$ to each $y \in \mathbb R$ can be computed as 
\begin{align*}
    \ell_{P_{XY}}(X \to y) &= \log \; \max_{x \in \{0,1\}} \; \frac{f_{Y \mid X} (y,x)}{f_Y(y)}\\
    &= \log \; \frac{2}{\exp \left(- {\textstyle \frac{\abs{y-\frac{1}{2}}}{\sigma^2}}\right) + 1}. 
\end{align*}
Specifically, $\ell_{P_{XY}}(X \to \frac{1}{2}) = 0$ and $\lim_{y \to \infty} \ell_{P_{XY}}(X \to y) = \lim_{y \to -\infty} \ell_{P_{XY}}(X \to y) = \log 2$.  
\end{example}

\begin{example}[Bivariate Gaussian]
\label{ex:gaus_2}
Suppose $X$ and $Y$ are zero-mean jointly Gaussian random variables with variances $\sigma_X^2$ and $\sigma_Y^2$, respectively, and correlation coefficient $\rho \in (-1,1)$. 
Then, $Y \mid X=x \sim \mathcal N({\textstyle\frac{\sigma_Y}{\sigma_X}} \rho x, (1-\rho^2) \sigma_Y^2))$, and the PML from $X$ to $y \in \mathbb R$ is 
\begin{equation*}
    \ell_{P_{XY}}(X \to y) = \begin{cases}
    \frac{y^2}{2 \sigma_Y^2} - \frac{1}{2} \log (1 - \rho^2) &  \mathrm{if} \; \rho \neq 0,\\
    0 & \mathrm{if} \; \rho = 0.\\
    \end{cases}
\end{equation*}
\end{example}


\end{document}